\documentclass[]{article}
\usepackage{graphicx}
\usepackage[colorlinks=true,breaklinks=true,citecolor=blue,linkcolor=blue,urlcolor=blue]{hyperref}

\title{{\Huge Avrocar: a real flying saucer}}

\author{\textbf{Desire Francine G. Fedrigo*}\\  Panoramic Residence, Rua Lu\'{i}sa, 388s, ap. 05,\\ Vila Portuguesa, Tangar\'{a} da Serra/MT, 78300-000, Brasil\\ 
\textbf{Ricardo Gobato}\\ Secretaria de Estado da Educa\c{c}\~ao do Paran\'{a} (SEED/PR),\\ Av. Maring\'{a}, 290, Jardim Dom Bosco,\\ Londrina/PR, 86060-000, Brasil\\
	 \textbf{Alekssander Gobato}\\ Faculdade Pit\'{a}goras Londrina, \\Rua Edwy Taques de Ara\'{u}jo, 1100,\\ Gleba Palhano, Londrina/PR, 86047-500, Brasil\\
	 *\textbf{*Corresponding author}: desirefg@bol.com.br}

\begin{document}

\maketitle
 

\begin{abstract}
 One of the most unusual military aircraft programs V / STOL was the Avro VZ-9 ``Avrocar''. Designed to be a real flying saucer, the Avrocar was one of the few V / STOL to be developed in complete secrecy. Despite significant changes in the design, during flight tests, the Avrocar was unable to achieve its objectives, and the program was eventually canceled after an expenditure of 10 million US dollars between 1954 and 1961. But the concept of a lift fan, driven by a turbojet engine is not dead, and lives today as a key component of Lockheed X-35 Joint Strike Fighter contender. Was held in a data research and information related to Avrocar project carried out during the Second World War, which was directly linked to advances in aircraft that were built after it, and correlate them with the turbo fan engines used today.
 \end{abstract}

\section{Introduction}
One of the most unusual military aircraft programs V / STOL was the Avro VZ-9 ``Avrocar" (Figure 1). Designed to be a real flying saucer, the Avrocar was one of the few V / STOL to be developed in complete secrecy. Despite significant changes in the design, during flight tests, the Avrocar was unable to achieve its objectives, and the program was eventually canceled after an expenditure of 10 million US dollars between 1954 and 1961. \cite{2015}
 
\begin{figure}
 	\begin{center}
 		\includegraphics[scale=0.65]{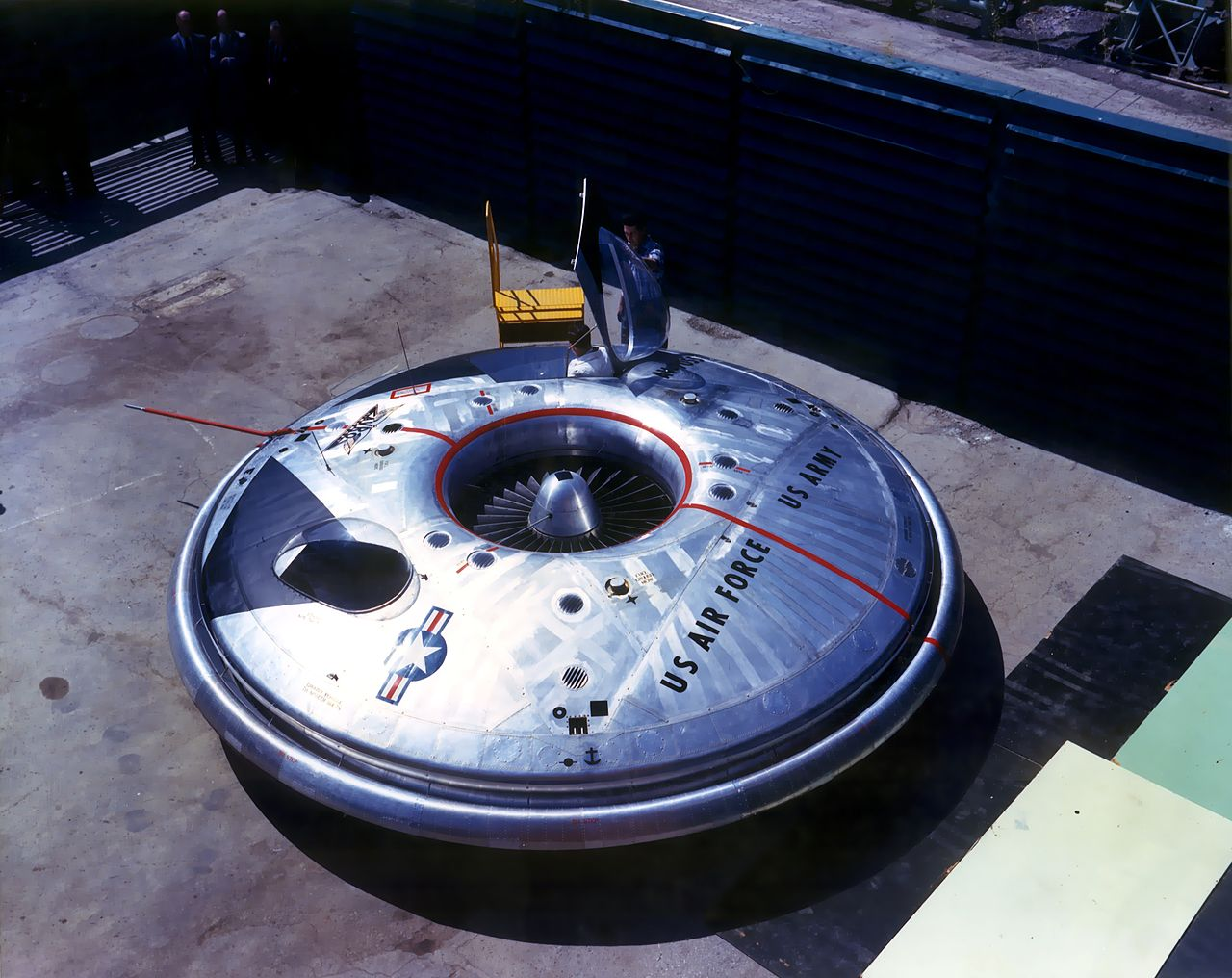}
 		\caption{\small{The Avrocar: authentic flying saucer. \cite{2015}}}\label{fig:sweeps}
 	\end{center}
\end{figure}
 
\subsection{Material and methods}
 
Based on books, documents available in digital media and in public libraries and using conventional qualitative research methods descriptive nature the work is presented. Noting also of the 40s books, related to the project Avrocar and their respective manufacturing, development of aircraft and engines that followed the same turbo fan concept prototype after World War II. 

\section{Project development Avrocar}
In 1952, a project team led by Jack Frost, the Avro Aircraft, Canada, began work on the design of an aircraft V / TOL supersonic with a circular wing. The Canadian Defense Research Board funded the project, with a contract of \$ 400,000. The Capacity V / TOL was to be done by the air ducts of the fan and the engine exhaust to the periphery of the platform, diverting the flow of air down. Near the ground, this provides an effect of ``air mattress", where the weight exceeds the pressure due to increased pressure on the underside of the aircraft. This phenomenon was confirmed in a wind tunnel test. In transit to forward flight, the airflow was gradually redistributed back. Frost was convinced that a circular wing thin disk, was the ideal format to take advantage of both the effect of ``air cushion" close to the ground (for STOL) and to achieve supersonic speeds.\

\begin{figure}
	\begin{center}
		\includegraphics[scale=0.4]{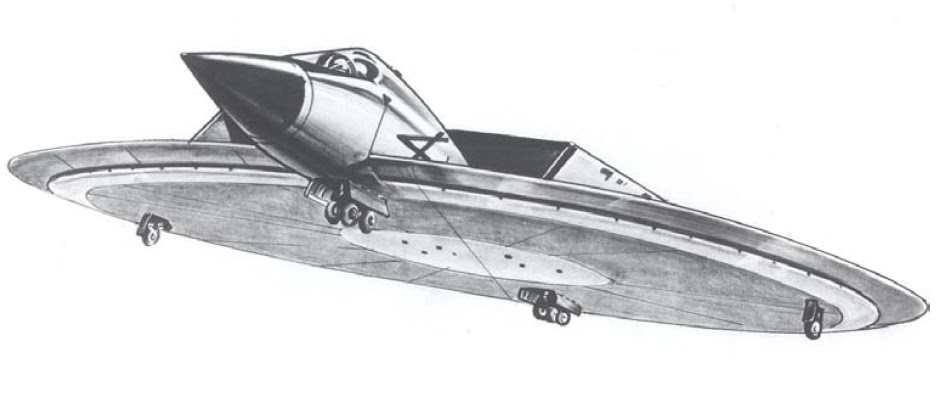}
		\caption{\small{Project 606A}. \cite{Gobato2011}}\label{fig:sweeps}
	\end{center}
\end{figure}

In 1954, the Canadian government abandoned the project to be very expensive, but the progress that had been made was enough to interest the United States Air Force (USAF). Concern about the vulnerability of air bases on the front line of Europe, the Cold War, has increased the interest of the Air Force V / TOL aircraft. For three quarters of a million dollars, the contract was signed by the Air Force in 1955 for further study. In 1956, the Avro was sufficiently satisfied with the results to release 2.5 million dollars to build a prototype of the search plane. In March 1957, the Air Force approved additional funding, and the aircraft became, officially, the ``System 606A Arms".\

These efforts remained highly classified as top secret until July 1960. One of the most promising proposals was the 606A (Figure 2) that would have a circular thin wing 35 meters in diameter, with a maximum weight of 27,000 pounds and a speed of more than Mach 1.4. A large turbofan engine was driven by the exhaust gas flow turbojets Armstrong Siddeley six Viper.\

Numerous wind tunnel testing in both the Avro as the USAF test base at Wright Field, Ohio, were conducted and a test platform, full-scale, the propulsion system was built. The Avrocar 59-4975 after modifications, was tested without the canopies and incorporating the perimeter ``focusing" ring c. 1961 (Figure 3). Tests showed that the heat was so oppressive that all instruments were baked brown after only a few flights (Figure 3).\

\begin{figure}
	\begin{center}
		\includegraphics[scale=1]{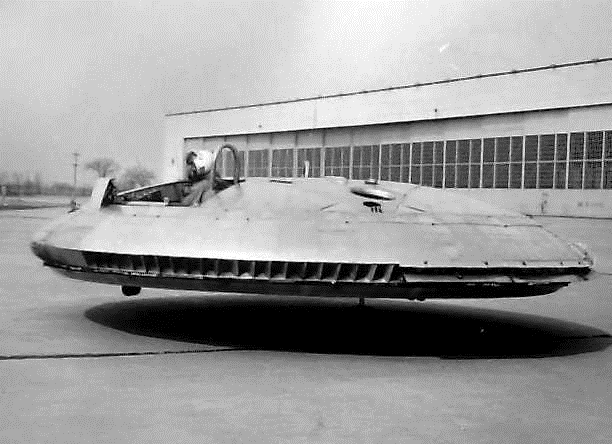}
		\caption{\small{Avrocar in flight in ground effect. Avrocar 59-4975 after modifications, was tested without the canopies and incorporating the perimeter ``focusing" ring c. 1961. Tests showed that the heat was so oppressive that all instruments were baked brown after only a few flights. \cite{Gobato2011, Force2009, 2009}}}\label{fig:sweeps}
	\end{center}
\end{figure}

In 1958, the Avro made a series of presentations to the Army and United States Air Force, after which the Avro began a project of an aircraft for the US military, which has been given the official designation VZ-9. It was baptized with the name of Avrocar (Figure 4 and Figure 5).

\begin{figure}
	\begin{center}
		\includegraphics[scale=0.6]{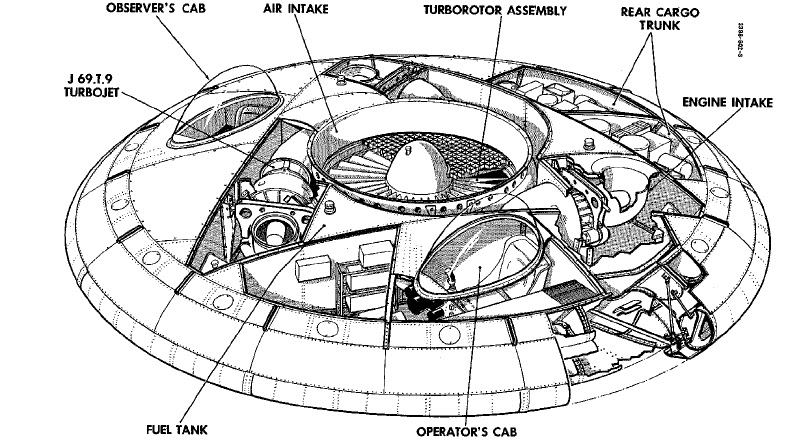}
		\caption{\small{VZ-9 Avrocar. \cite{2015, Campagn2003}}}\label{fig:sweeps}
	\end{center}
\end{figure}

\begin{figure}
	\begin{center}
		\includegraphics[scale=0.3]{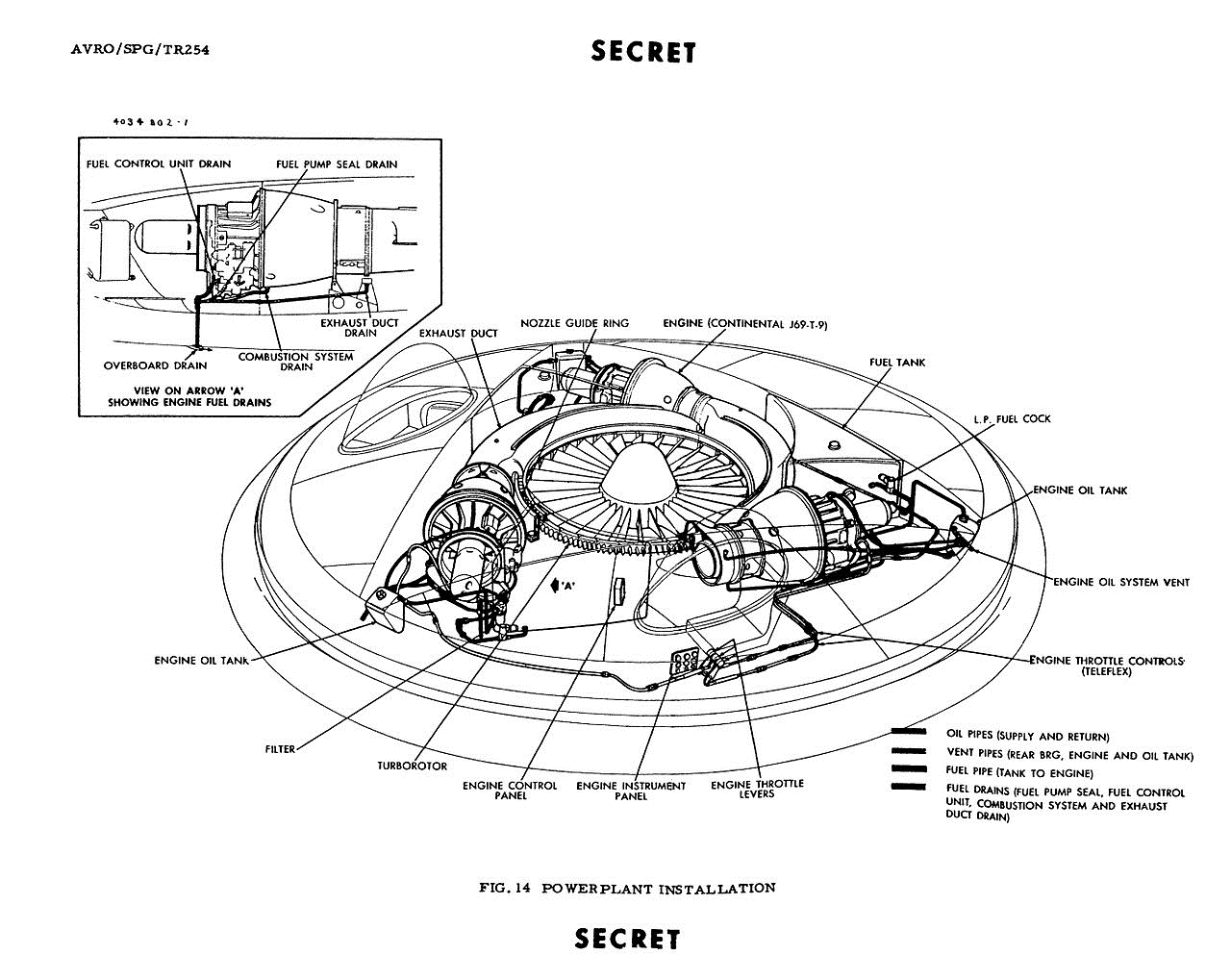}
		\caption{\small{VZ-9 Avrocar. \cite{2015, Campagn2003}}}\label{fig:sweeps}
	\end{center}
\end{figure}

The Avrocar (Figures 7-9) should be a circular platform aircraft, which to present both hovered flight and horizontal flight at high speed, with capacity V / TOL. The Army was interested in the survival of troops on the battlefield and in order to improve its air capacity, they were studying alternatives to their aircraft and helicopters then in service.\

The Air Force supported the Avrocar program because it would demonstrate many of the features of the 606 project, in less time and at a much lower cost. A two million dollar contract, to be managed by the Air Force was made in the Avro to build and test an Avrocar.\

The initial performance requirements for Avrocar were: be able to sustain ten minutes flight in ground effect and carry a load of 1000 pounds, 25 miles away.\

The work began in earnest, and a contract 1.77 million dollars was granted to build a second copy of Avrocar in March 1959. The first Avrocar left the factory in May 1959. At launch, the projected performance was very beyond the initial requirement, and reach a top speed of 225 kt, ceiling 10,000 feet, 130 miles of range with 1,000 kg payload, and hover out of ground effect, with 2428 kg of payload. The maximum take off weight, with transition to forward flight, out of ground effect, it was estimated at 5,650 pounds and the maximum weight, with a transition in ground effect, it would be 6,970 pounds.\

The Avrocar was about 18 meters in diameter, three meters of disc thickness, and two separate cockpits. The cockpit was located on the left front side of the aircraft, with another crew member on the right. A third compartment on the back was due to charge storage. The Avrocar would be raised by the flow of a turbo-fan engine five meters core diameter, called exhaust turbo-rotor, composed of three Continental J-69 turbojet with 920 pounds of thrust each, whose air flow was channeled to the outer edge of the turbo rotor. It had 124 small straws. Each engine was connected to its own fuel tanks and oil. The fuel tanks were not connected, although it had planned for a later version.\ 

The main pilot control consisted of a side-stick, which provided control pitching and side when moved forwards and backwards or sideways. The yaw control was obtained twisting the side-stick. This action controlled low flows and high pressure around the aircraft, causing it to rotate around the vertical axis.\

In forward flight, the Avrocar was statically unstable, with a pressure center far ahead of the center of gravity. An automatic stabilization system was employed, using the gyroscopic action of the turbo-rotor. The turbo-rotor, in turn, was not rigidly attached to the vehicle, but mounted on a kind of hinge that allowed him freedom of movement. Control cables are set on the basis of the turbo rotor, to enable its control.\

The interest of the Army in Avrocar program was great. One of the authors surveyed (Lindenbaum) recalls a trip that made Washington in the late 1950s, to request additional funding for a study on reducing the aerodynamic drag of helicopter Bell UH-1. Although funding had been approved, he heard a note of an army general, that the Bell UH-1 Huey helicopter would be the last that the army would buy, since the helicopter would be replaced by Avrocar.\

From June to October 1959, the first Avrocar was tested on a static platform, hovering. The hot gases re-circulated the turbo-rotor reduced the thrust. Excessive losses in the duct system also became apparent, and these defects have never been solved, despite major changes in the design. The maximum load reached, out of ground effect, was 3,150 pounds. With a zero fuel weight (ZFW) of 4285 pounds, the Avrocar was therefore unable to hover out of ground effect. Following these tests, the vehicle was sent to NASA for assessing wind tunnel Ames Research Center.\

The second Avrocar left the factory in August 1959. On September 29, the first attempt of sustained flight was made with the Avrocar pinned to the ground by cables. After that the vehicle took off, an uncontrollable oscillation occurred with each wheel alternately bouncing on the ground. The pilot immediately shut down all engines. Subsequently, several alternative schemes, these ``captive" flights have been tested and numerous changes were made to the springs of spoilers and control of rotor shaft base. These first captive flight revealed then a new problem, called ``hubcapping" which has never been fully resolved. The hubcapping was rapid and unpredictable swings in pitch and roll axes. It resulted in an unstable air mattress if the vehicle exceeds a critical height (Figure 6). \cite{Force2009, Campagn1998, DesireFrancineGobato2011, Stevens2003}\

\begin{figure}
	\begin{center}
		\includegraphics[scale=0.8]{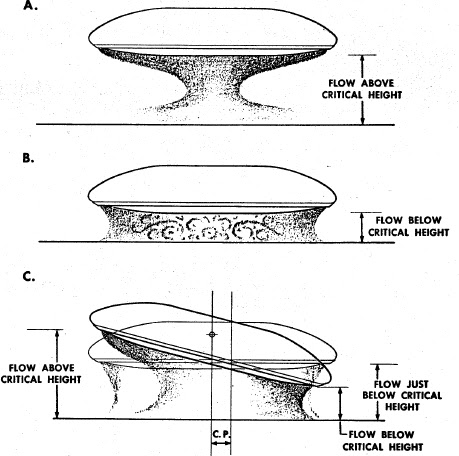}
		\caption{\small{Critical altitude of Avrocar, and the hubcapping. \cite{Zuk2002, Zuk2006}}}\label{fig:sweeps}
\end{center}
\end{figure}

The critical point was found about two meters from the ground. Inputs the controls were ineffective in dampening the oscillation. Fifty-two holes were drilled at the vehicle bottom, located radially and three meters from the center. These were used to provide a flow of air attempting to center and stabilize the air mattress. Such a device never reached the expected success.\

\begin{figure}
	\begin{center}
		\includegraphics[scale=0.65]{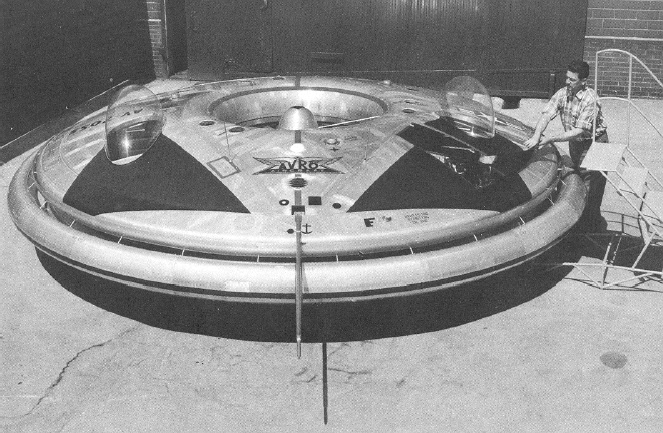}
		\caption{\small{VZ-9 Avrocar. \cite{Zuk2002, Zuk2006, 2015a}}}\label{fig:sweeps}
	\end{center}
\end{figure}

\begin{figure}
	\begin{center}
		\includegraphics[scale=1.3]{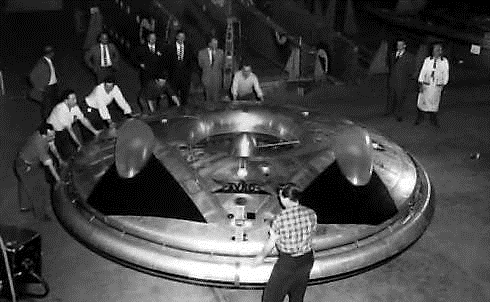}
		\caption{\small{VZ-9 Avrocar. \cite{Zuk2002, Zuk2006, 2015a}}}\label{fig:sweeps}
	\end{center}
\end{figure}

\begin{figure}
	\begin{center}
		\includegraphics[scale=0.85]{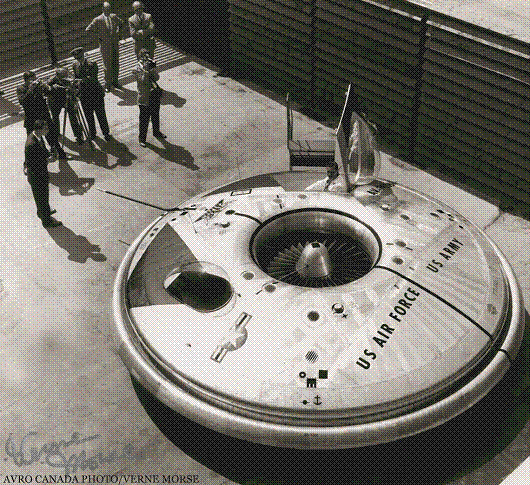}
		\caption{\small{AVRO Canada, US ARMY, US AIR FORCE, Photo/Verne Morse. \cite{Zuk2002, Zuk2006, 2015a}}}\label{fig:sweeps}
	\end{center}
\end{figure}

The first completely free flight occurred on November 12, 1959, and the control system based on adjustable nozzle air outlet, proved unacceptable. After five flights, the test was temporarily interrupted in December 5, 1959, when the Avrocar had recorded 18.5 hours of testing in captive and free flights.\

A new control system, centered on a set in a ring, was installed later in December. The tip opening to the upper surface was covered, and spoilers were replaced by a flat ring the underside of the vehicle. Lateral changes in position of the ring increased the weight on one side of the vehicle while reducing weight of the opposite side.\

Flight tests resumed in January 1960, with this system. The trial flight of the Air Force was held on April 4, 1960, with Major Walter Hodgson controls. The maximum speed reached was 30 Kts, and above this speed, an uncontrollable oscillation manifested. The cockpit was cramped, noisy and became unbearably hot during a flight of 15 minutes. Later that month, a trial was conducted in a wind tunnel at NASA's Ames Research Center in Moffet Field.\

This test discovered that the control ring centered system provided sufficient buoyancy to allow a flight out of the ground effect, but large angles of attack were needed to generate aerodynamic lift. In late April, however, the Avrocar the initial program ended. Shortly after the program was declassified by the USAF HQ.\

The Avro was convinced that the concept was still viable, and proposed a new program to rework the main failures of propulsion and control system (Figure 10).\

\begin{figure}
	\begin{center}
		\includegraphics[scale=0.6]{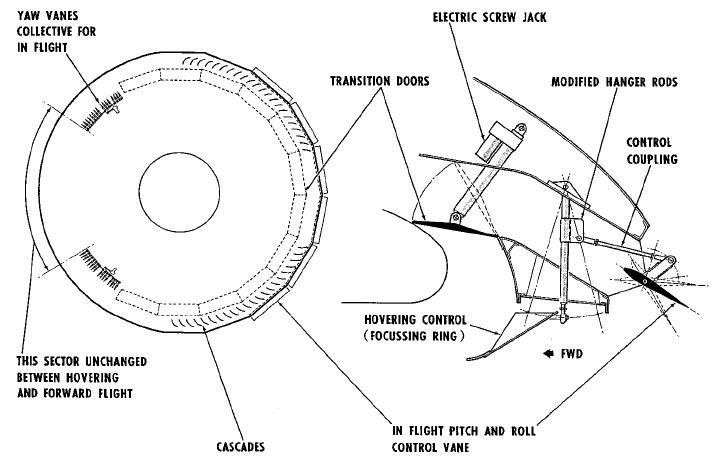}
		\caption{\small{High speed control system. \cite{Gobato2011, DesireFrancineGobato2011, Zuk2002,  Gobato2011a}}}\label{fig:sweeps}
	\end{center}
\end{figure}
The USAF made a new contract for the period July 1960 to July 1961 for the modification and testing of both vehicles. A new control nozzle is installed at the rear of the vehicle.\ 

A second test in the wind tunnel with the new configuration was done at NASA Ames RC in April 1961. It has been found that an adequate control was available for the transition to a speed of about 100 Kts, and the flight was possible at this speed. However, the vehicle was still unstable. It was hoped that the change of flow over the rear of the vehicle lift the nose up, reducing the instability. Unfortunately, this was not the case.\

A tail ``T" has been added (Figure 11), but it has proved totally ineffective. NASA believes that this failure resulted from the fact that the tail was in a region of ``downwash" too high caused by the propulsion system. In any case, it became clear that the Avrocar as configured, could not sustain flight high speed.\

\begin{figure}
	\begin{center}
		\includegraphics[scale=0.7]{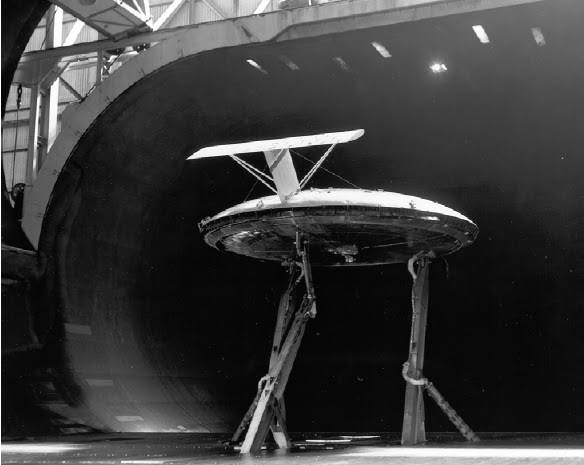}
		\caption{\small{The Avrocar with a tail added on T. \cite{Gobato2011, Zuk2006}}}\label{fig:sweeps}
	\end{center}
\end{figure}

On June 9, 1961, the final evaluation and second flight of the Air Force Avrocar was conducted in Avro facilities. During these tests, the vehicle reached a top speed of 20 Kts and showed the ability to cross a ditch six meters wide and 18 centimeters deep. Fly above the critical time was impossible. The flight test report summarized a number of control problems. For example, a large asymmetry in directional control was present. Five seconds were required to turn the aircraft 90 degrees to the left, while eleven seconds were required for a right turn of 90 degrees.\
  
\begin{figure}
	\begin{center}
		\includegraphics[scale=0.7]{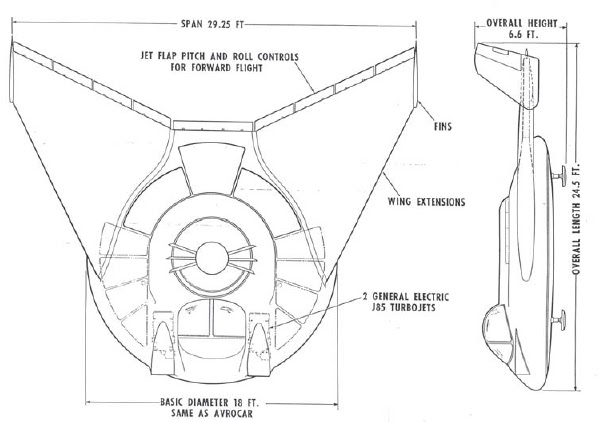}
		\caption{\small{Wings and winglets added to Avrocar. \cite{Force2009, Rose2007}}}\label{fig:sweeps}
	\end{center}
\end{figure}

The Avro submitted proposals for radical change in vehicle to address the key problems. Frost (team member) has developed two new models, one with a large vertical tail and another with winglets vertical (Figure 12). Both models would use two turbojets GE-85 J 2700 pounds of thrust, rather than the three original J-69 turbines and the turbo rotor diameter would be increased from five to six feet. The proposals were rejected, and the program was officially closed in December 1961. The second Avrocar had recorded about 75 hours of flight. The Figure 13, shows the Avrocar at the National Museum of the United States Air Force in Dayton, Ohio. \cite{Gobato2011, Force2009, Campagn2003, Stevens2003, Gobato2011a, Zuk2002, Zuk2006}\

\begin{figure}
	\begin{center}
		\includegraphics[scale=0.85]{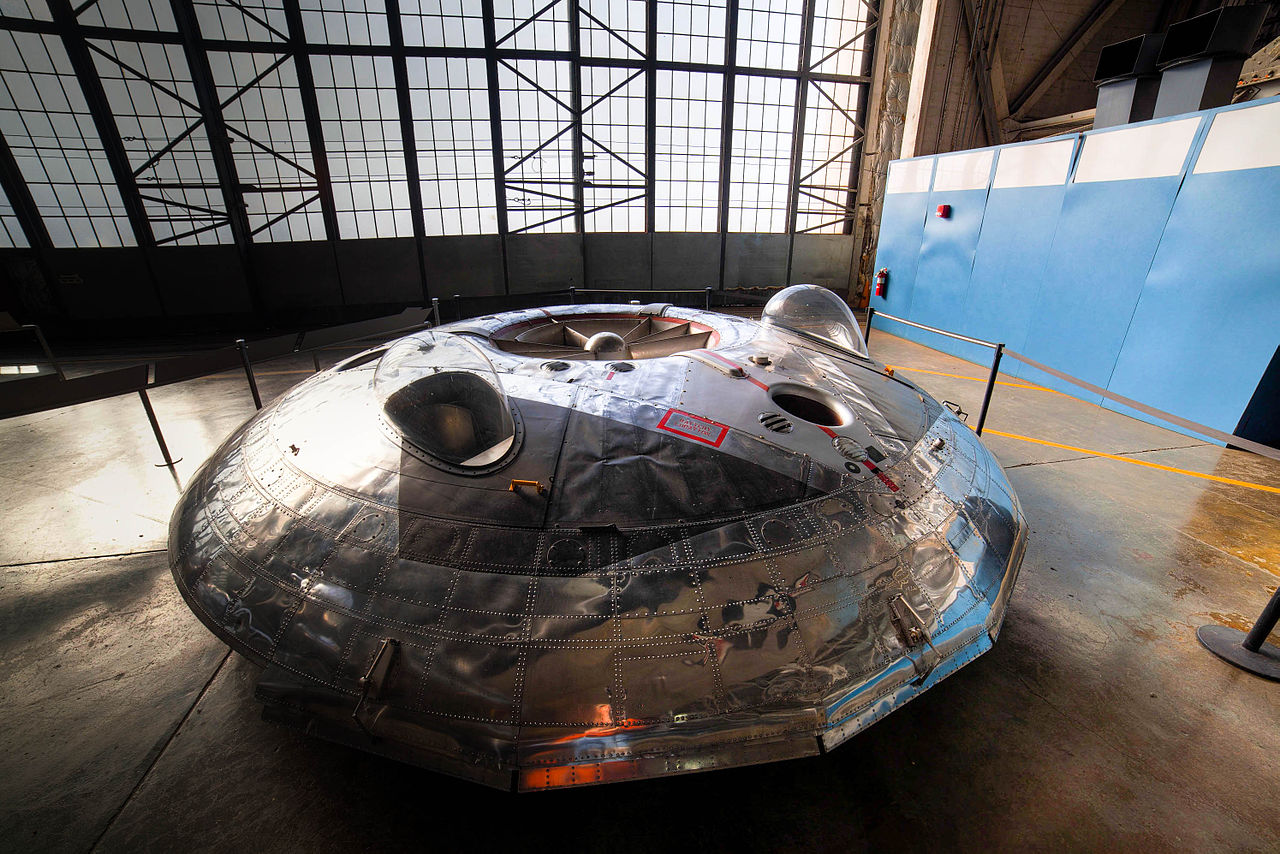}
		\caption{\small{Avrocar at the National Museum of the United States Air Force in Dayton, Ohio. \cite{2015a}}}\label{fig:sweeps}
	\end{center}
\end{figure}

\section{Discussions}
 The concept of ground effect produced by a fan at takeoff and landing did not die with the Avrocar. In 1963, Bell Aerospace initiated studies of a landing system for air mattress (ACLS), which was later patented. These studies were directed by Desmond T. Conde, former chief of aerodynamics for Avrocar.\
 
 A ACLS replace conventional landing gear with a large internal rubber tube-like structure which surrounds a region of higher air pressure. In August 1967, the concept was proved by Bell, with successful tests on a LA-4 (amphibian). The development was funded by the Air Force Flight Dynamics Laboratory, and a much larger system is designed to test on a Fairchild C-119 (weight of 64,000 lb).\ 
 
 The Brazilian government joined the program and a De Havilland C-115 Buffalo aircraft weighing 41,000 pounds, was selected for further testing. With the XC-8A designation, this aircraft flew with the ACLS in March 1975. The ACLS was considered but rejected as an option for the program from a STOL aircraft / Midfielder Forward, a program that ended up producing the Boeing YC-prototypes 14 and McDonnell Douglas YC-15. The latter was adopted and evolved into the Boeing C-17 transport, which went into series production.\
 
 The concept of a lift fan, driven by a turbojet engine is not dead, and lives today as a key component of Lockheed X-35 Joint Strike Fighter contender. While Avrocar was in development, Peter Kappus, General Electric has developed independently by a booster fan propulsion system, which has evolved to Ryan VZ GE-11 (later XV-5) ``Vertifan". This vehicle, discussed in the previous two editions of Vertiflite magazine (March / April 1990 March / April 1996), paved the way for further study of the ``fans elevators", or supportive fans, as the study of supersonic fighters sponsored by DARPA which included both fans driven gas (McDonnell Douglas) and mechanically driven shaft (Lockheed).\

\section{Conclusions} 
Through research carried out during the work it was found that despite the possible failure Avrocar the project was responsible for the breakthrough in the aeronautical environment and their van turbo engines, in which the project was considered essential to the development and improvement of existing fan motors. It is hoped that with this work can be broken the view that many drivers have about the project, which is considered as a bad investment of time since it is not succeeded in the course of its development but it sure your project was in history and served as a study base for future projects.\

The concept of ground effect produced by a fan at takeoff and landing did not die with the Avrocar. In 1963, Bell Aerospace initiated studies of a landing system for air mattress (ACLS), which was later patented. These studies were directed by the former head of Avrocar project, Desmond T. Conde. The Avrocar also ended up producing the prototype Boeing YC-14 and McDonnell Douglas YC-15. The latter was adopted and evolved into the Boeing C-17 transport, which went into series production. This concept lives today as a key component of Lockheed X-35 Joint Strike Fighter contender. While Avrocar was in development, Peter Kappus, General Electric has developed independently by a booster fan propulsion system, which has evolved to Ryan VZ GE-11 (later XV-5) ``Vertifan".\

\bibliographystyle{unsrt}
\bibliography{journals}

\end{document}